\documentclass[a4paper,nofootinbib,floatfix]{article}
\pdfoutput=1 
             
\usepackage{graphics}
\usepackage{graphicx}
\usepackage{epsfig}
\usepackage{amssymb}
\usepackage{amsmath}
\usepackage{jcappub} 

\title{\boldmath Pinpointing astrophysical bursts of low-energy neutrinos embedded into the noise}

\author[a,b]{C. Casentini,}
\affiliation[a]{Universit\`a di Roma Tor Vergata, via della Ricerca Scientifica, I-00133 Roma, Italy}
\affiliation[b]{INFN Sezione di Roma Tor Vergata, via della Ricerca Scientifica, I-00133 Roma, Italy}
\author[c,d,1]{G. Pagliaroli\note{Corresponding author},}
\affiliation[c]{Gran Sasso Science Institute, Viale Francesco Crispi 7, I-67100 L'Aquila, Italy}
\affiliation[d]{INFN-LNGS, I-67100 L'Aquila, Italy}
\author[e]{C. Vigorito,}
\affiliation[e]{Universit\`a di Torino $\&$ INFN, via Pietro Giuria 1, I-10125 Torino}
\author[a,b,c]{V. Fafone}

\abstract{
We propose a novel method to increase the probability of identifying impulsive 
astrophysical bursts of low-energy neutrinos. The proposed approach exploits 
the temporal structure differences between astrophysical bursts and background fluctuations 
and it allows us to pinpoint weak signals otherwise unlikely to be detected. 
With respect to previous search strategies, this method strongly reduces
the misidentification probability, e.g. for Super Kamiokande this reduction is a 
factor of $\sim 9$ within a distance of $D\sim 200$ kpc without decreasing the detection efficiency. 
In addition, we extend the proposed method to a network of different detectors and 
we show that the Kamland $\&$ LVD background reduction is improved 
by a factor $\sim 20$ up to an horizon of $D\sim75$ kpc. 
}

\begin{document}
\maketitle

\section{Introduction}
Core-Collapse Supernovae (CCSNe) \cite{Janka:2006fh} represent the final explosive phase of massive stars and
the detection of a galactic event could be the unique opportunity for us to grasp the physical mechanism driving 
the final explosion of the structure. 
``Failed" Supernovae \cite{Adams:2016hit} are collapsing stars failing to explode and 
forming an inner black hole. The lack of the final explosion makes these sources optically silent and, 
at the present, have never been directly observed. 
Quark Novae \cite{Ouyed:2001ts} are expected when a neutron star suddenly converts into a quark star. Their 
existence is strongly related to the fundamental state of the matter  and their detection could provide 
the first clear evidence of the presence of strange matter in the universe. 

A common signature for all these catastrophic astrophysical phenomena is expected to be an impulsive 
$(\sim 10$ s$)$ emission of low-energy, $(\sim 10$ MeV$)$, neutrinos \cite{Totani:1997vj,Sumiyoshi:2007pp,Keranen:2004vj}. 
Despite the large amount of total energy $(\sim10^{53} \text{ergs})$ released in neutrinos, when the source distance
increases and/or the average energy of emitted neutrinos decreases, the signal statistics drops and 
the identification of these astrophysical bursts embedded into the detector noise could be challenging. 

The search of these astrophysical signals is one of the main goals of several low-energy neutrinos  
detectors based on different techniques and characterised by different capabilities (see \cite{Scholberg:2012id} for a Review).  
Moreover, the SuperNova Early Warning System (SNEWS) \cite{SNEWS} provides an early warning of a galactic supernova
demanding the coincident observation of low-energy neutrinos bursts from several detectors. 
The detection strategy adopted in each neutrino detector, working alone or in the SNEWS system, 
relies on the competition between the background rate collected in a 
fixed time window and the expected signal rate in the same time interval. 
In order to separate real signals from background fluctuations, the standard procedure
is purely statistical, data are selected requiring that the Poisson probability that background 
fluctuation produces the observed number of events is very small. 
This means that only very strong signals are well identified whereas
small signals cannot be separated from statistical fluctuations and are unavoidably lost. 

In this paper, we improve the detectors capability to disentangle astrophysical bursts of low-energy 
neutrinos from background signals. This powerful method exploits the different temporal structure 
expected for an astrophysical burst with respect to background fluctuations that are near uniformly distributed in a time window.
This characteristic, described with a new parameter, can be used as an additional degree of freedom that, added to the 
statistical requirement, improves our capability to identify real signals allowing the detection of weaker/far away 
astrophysical sources.  

\section{Assumptions}
For all the astrophysical sources we are interested in, we assume 
that the total energy radiated in neutrinos is $\mathcal{E}=3\cdot 10^{53}$ erg.
Moreover, based on CCSN study, we consider that the total energy is 
partitioned in equal amount among the six types of neutrinos, 
that should be true within a factor of 2 \cite{Keil:2002in}.

As highlighted, the novelty of the proposed method
is the introduction of a discrimination parameter based on the different temporal shapes
of background and signal. We consider a very general description of an astrophysical burst of low-energy neutrinos 
characterised by the following temporal evolution
\begin{equation}
f(t)=[1-\exp(-t/\tau_1)]\exp(-t/\tau_2), 
\label{tempo}
\end{equation}
where $\tau_1 =(10-100)$ms is the rising time and $\tau_2\ge1$ s 
represents the decaying time of the signal. 
This ansatz is very general, model independent and it fits all the expected neutrino 
emissions from CCSNe\cite{Totani:1997vj}, Failed Supernovae\cite{Sumiyoshi:2007pp} and 
Quark Novae\cite{Ouyed:2001ts}. 
Moreover, SN1987A, the only CCSN detected 
so far by neutrinos telescopes, agrees with this simple temporal model. The complete set 
of events observed by Kamiokande-II\cite{KII}, IMB\cite{IMB} and Baksan\cite{Baksan} fitted with 
such a model provides a best-fit time scale of $\tau_2\sim 1$s \cite{noi}.   

For the energy differential fluence we assume quasi-thermal spectra described by
\begin{equation}
\Phi_i^0=\frac{\mathcal{E}_i}{4\pi D^2}\times \frac{E^\alpha e^{-E/T_i}}{ T_i^{\alpha+2} \Gamma(\alpha+2)} \ \ \
  i=\nu_e,\nu_\mu,\nu_\tau,\bar{\nu}_e,\bar{\nu}_\mu,\bar{\nu}_\tau,
\end{equation}
where $E$ is the neutrino energy, $\mathcal{E}_i=\mathcal{E}/6$ is the energy radiated in each specie 
due to the equipartition hypothesis, 
and the `temperature' is $T_i=\langle E_i\rangle/(\alpha+1)$.
The average energy per flavour is $\langle E_i\rangle$ and the parameter 
$\alpha=3$ represents a mild deviation from a thermal distribution.
According to SN1987A data\cite{noi} and recent numerical simulations \cite{Tamborra:2017ubu},
we set $\langle E_{\nu_e}\rangle=9$ MeV,  $\langle E_{\bar{\nu}_e}\rangle=12$ MeV and non-electronic temperature,
$\langle E_{\nu_x}\rangle$, 30\% higher than $\langle E_{\bar{\nu}_e}\rangle$.   

Here, we consider the main interaction channel, namely 
the inverse beta decay (IBD) $\bar\nu_e+p \rightarrow n+e^+$.
Due to neutrino oscillations, the $\bar{\nu}_e$ fluence at the detector 
is an admixture of the unoscillated flavour fluences at the source: 
$\Phi_{\bar{\nu}_e}=P \Phi^0_{\bar{\nu}_e}+ (1-P)\Phi^0_{x}$, where $x$ indicates 
the non-electronic flavours and $P$ is the survival probability for the $\bar{\nu}_e$. 
Depending on the neutrinos mass hierarchy, this probability can be $P=0$ for Inverted Hierarchy (IH) 
or $P\simeq 0.7$ for Normal Hierarchy (NH). 
The expected number of IBD interactions is $S(E_{\nu},D)= N_p\sigma_{\bar\nu_e p}(E_{\nu})
\Phi_{\bar\nu_e}(E_{\nu},D)\epsilon (E_{vis})$, where $D$ is
the source distance, $N_p$ is the number of target protons within the
detector, $\sigma_{\bar\nu_e p}$\cite{strum} is the process cross section and
$\epsilon$ is the detector efficiency as a function of the visible energy $E_{vis}$. 
For an energy threshold $E_{thr}\sim 1$ MeV, the positron energy spectrum of the IBD channel 
is completely observed and, in the optimistic case of a total detection efficiency ($\epsilon=100\%$), 
a fixed number of IBD interactions $S$ can be obtained in three different cases: 
1) Inverted neutrino mass hierarchy, average energy $\langle E_{\bar{\nu}_e}\rangle$=$12$ MeV 
and source distance $D$; 
2) Normal neutrino mass hierarchy, average energy $\langle E_{\bar{\nu}_e}\rangle$=$12$ MeV 
and source distance $0.896 D$; 
3) NH, $\langle E_{\bar{\nu}_e}\rangle$=$15$ MeV and source distance $D$.  
In other words, the effect on the number of events due to 
a change of the source distance or of the average energy of the spectrum 
or of the neutrino mass hierarchy is the same. 
By taking into account this degeneracy we show our results for the 
NH case with $\langle E_{\bar{\nu}_e}\rangle$=$15$ MeV. 
By using previous considerations these results can be 
rescaled to the IH case or to a different neutrino average energy.

\section{Method}
The aim of this paper is to show an efficient method to discriminate astrophysical burst of low-energy neutrinos
from fake burst of events induced by background. For this reason background knowledge and characterization
is fundamental to demonstrate the potential of this method. At the present, low-energy neutrinos detectors 
on data-taking, viz. Super Kamiokande\cite{Abe:2016waf}, LVD\cite{Agafonova:2014leu}, Borexino\cite{2009BorexinoColl} and KamLAND\cite{2003KamlandColl} provide all the information needed to perform this study and we report results 
for these detectors considering both the situation where the detector operates alone and the case in which multiple detectors  
operate as a network exploiting the advantages of a combined coincident search.

In all the considered detectors the search of astrophysical bursts of low-energy $\nu$ is based  
on the definition of clusters of events.
Following \cite{Agafonova:2007hn} we define a cluster as the group of the events contained 
in consecutive time windows of $w = 20$ seconds. Each cluster is characterised by its multiplicity $m_i$,
i.e. number of events inside the time window, 
and its time duration $\Delta t_i$, defined as the time difference among
the first and the last event detected.
In order to increase the detection probability this search is performed one more time 
by shifting the consecutive time windows of $10$ seconds 
with respect to the first search (for more details see \cite{Agafonova:2007hn}).

In order to claim the detection of an astrophysical burst two different requirements should be satisfied:
the cluster of expected events for a specific source distance should be populated enough (at least two neutrinos $m_i\ge 2$) 
and this cluster should be discriminated by the others due to standard background events. 
    
In the following, the first requirement will be discussed in term of detection efficiency, $\eta$, and the second one will 
be related to the misidentification probability $\zeta$, i.e. the probability to confuse a background cluster for
a signal cluster. Both these quantities are strongly related to the background characteristics of each neutrino 
detector. In order to reproduce the background fluctuations of each detector and according to the parameters 
(frequencies and energy thresholds) reported in Tab.\ref{tab:freq}, a Monte Carlo simulation of $10$ years of data-taking has been performed .
For the observed background clusters of events we calculate its imitation 
frequency $f^{im}_i (\text{day}^{-1})$, i.e. how many times in a day 
background events produce a cluster with the same multiplicity. 
This quantity is defined as:
\begin{equation}
f^{im}_i=N\cdot\sum_{k=m_i}^\infty \frac{(f_{bkg} w)^k e^{-f_{bkg} w}}{k!} \text{ day}^{-1},
\end{equation}
where $N=8640$ is the number of windows of duration $w=20$ seconds overlapped 
every 10 seconds in a day, $m_i$ is the cluster multiplicity and $f_{bkg}$ is the background 
frequency of the experiment. In order to reduce the background, standard search 
procedures assume a selective cut on $f^{im}_i$ (or equivalently on $m_i$)\cite{Abe:2016waf, Agafonova:2014leu}.
Lower is the $f^{im}_i$ threshold value used higher is the probability that survived clusters are due to 
real astrophysical signals. For example, to reduce the background fluctuations to a negligible value,
the SNEWS threshold is $f^{im}\leq 1/{100\text{years}}$ \cite{SNEWS}. 
On the other hand, higher is the allowed value for the $f^{im}_i$ 
larger is the distance reach of our search or the sensitivity to weaker signals, 
since, here, the real signals with small statistics can enter in our analysis. 
To test the new method to discriminate signal from background, we set 
$f^{im}\leq 1/\text{day}$ as working threshold for this statistical cut. 

As discussed before, this background reduction is purely statistical and no physical characteristics 
of the signal are used in order to separate background clusters from signal clusters.
To perform this step forward, we simulate the signals expected in each neutrino detector by 
considering different source distances $D$ in the range $8.5-500$ kpc. 
Simulated signals are randomly injected inside the background. 
Once clusters are obtained following the previous procedure, we only select clusters 
with $f^{im}< 1/\text{day}$. 

\begin{figure}[h]
$$
\includegraphics[width=0.8\textwidth]{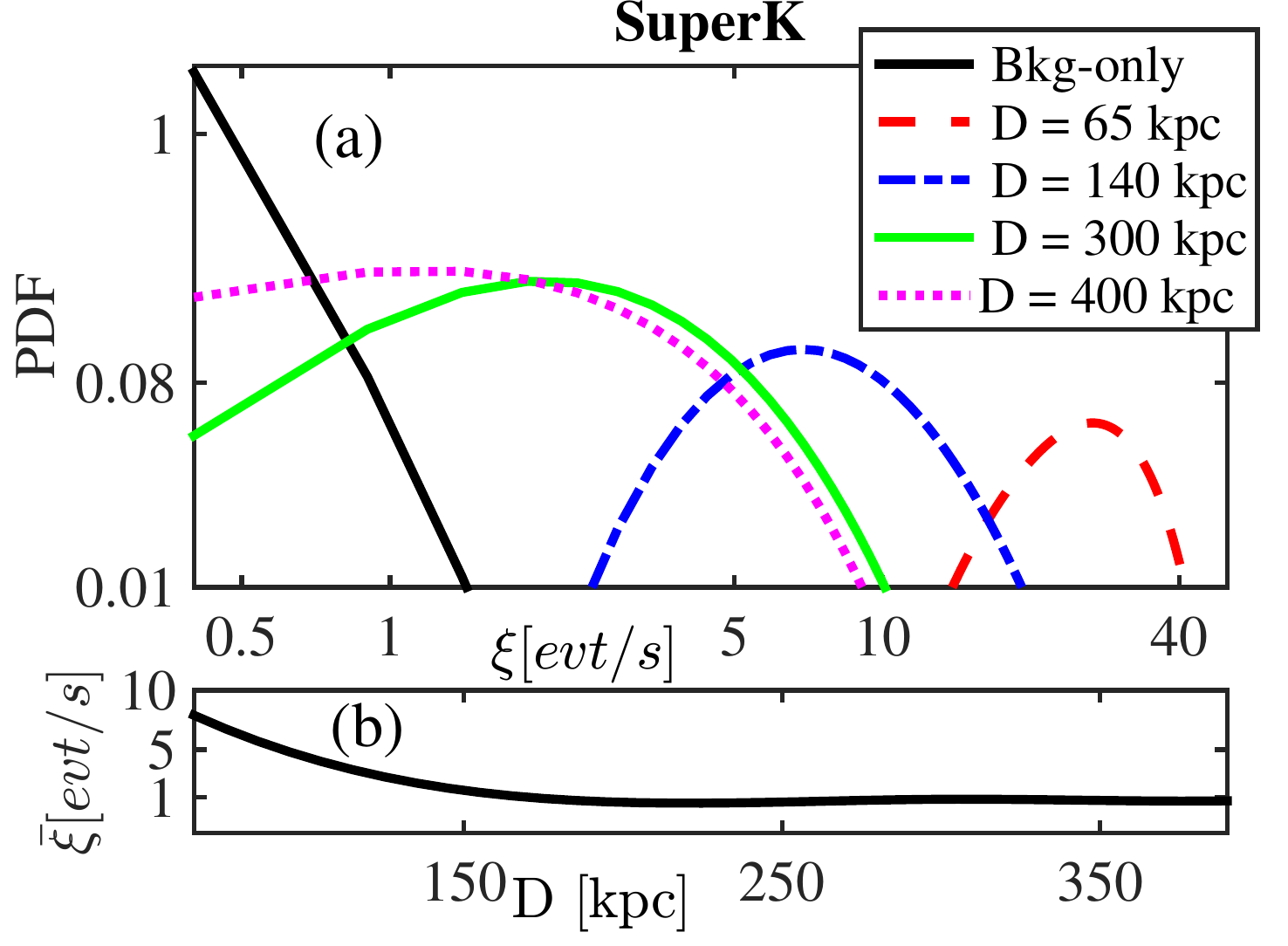}
$$
\caption{\em Panel (a): Probability density functions for background plus signal clusters 
as functions of the $\xi$ parameter and for different distances 
in the case of SuperK detector. The black solid line 
shows the PDF for pure background clusters; 
Panel (b): The optimal cut value for the $\xi$ parameter, $\overline{\xi}(D)$, as a function of the source distance D for SuperK detector. 
\label{fig1}}
\end{figure}

For each cluster we define the parameter, $\xi_i$ as the ratio between the cluster multiplicity and the cluster duration:
\begin{equation}
\xi_i={{m_i}\over{\Delta t_i}}
\label{xi}
\end{equation} 
and we study the $\xi_i$ distributions of pure background clusters 
and background plus signal clusters in term of normalised Probability Density Functions (PDF). 
In Fig \ref{fig1}(a) we show the result obtained for SuperK detector. 
All the PDFs are well described by a 4 parameters Gamma distribution. 
The distribution of clusters due to pure background events is reported with a black solid line and 
is characterised by very small values of the $\xi_i$ parameter as expected for events with a temporal 
uniform distribution, i.e. $\Delta t_i\rightarrow w$ seconds. 
On the other hand, clusters where also an astrophysical signal is present show a PDF shifted at higher values of $\xi_i$, 
as expected for clusters with events that cumulate faster in time $\Delta t_i < 20$ seconds. 
For any fixed source distance, the PDF is different and in Fig.\ref{fig1}(a) we show with different color
the case of D=65, 140, 300 and 400 kpc as expected in SuperK. Obviously, for closer source distances the 
expected multiplicity in presence of signal increases, allowing a better separation of the PDFs where astrophysical events are present. 
This separation becomes less and less as the source distance increases and the statistics decreases. 
To disentangle as well as possible the signal from the noise we define for each detector $X$ the function
\begin{align}
\notag{}\Xi[\xi]_X={\int_{0}^{\overline{\xi}_X} \text{PDF}^{bkg}_X\, d\xi}+{\int_{\overline{\xi}_X}^{\infty} \text{PDF}^{sig+bkg}_X(D)\, d\xi}
\label{minimizingnoise}
\end{align}
and we look for the $\overline{\xi}$ maximising this function. 
This value defines the best separation, for each detector $X$, between pure background distribution and signal plus background distribution. For any fixed source distance the PDF is different and a different optimal cut value for the $\xi$ parameter
can be defined. By performing several simulations, we define the function $\overline{\xi}(D)$, reported in Fig.\ref{fig1}(b) for the case of SuperK. 
As expected, we found a descendent behavior of the cut value with the increasing of the distance and this is observed for all the considered detectors. 
If the source distance is known, the optimal cut value of the new parameter is determined by this curve, however when a blind search of astrophysical signals is performed on real data the distance of the source is unknown. In this case we believe that the search 
should be optimised to the larger distance achievable. So that we define as optimal value of the cut parameter $\overline{\xi}_X$ the smaller one allowing a clear separation between background and signal PDFs. Finally, in the last column of Tab.\ref{tab:freq}, we report the optimal cut parameters found for each detector considered. 

As a consequence, we add as a new cut, on the statistically selected clusters, 
the condition $\xi_i \geq \overline{\xi}_X$ and we investigate 
its impact on the detection probability $\eta$ and on the misidentification probability $\zeta$. 
The detection probability is defined as the ratio between the number of signal clusters 
surviving after the cuts and the number of signal clusters initially injected into the background. 
In a similar manner the misidentification probability $\zeta$ is obtained as the fraction of background 
clusters that survive all the cuts over the total number of clusters observed. 

\begin{figure}[h]
$$
\includegraphics[width=0.8\textwidth]{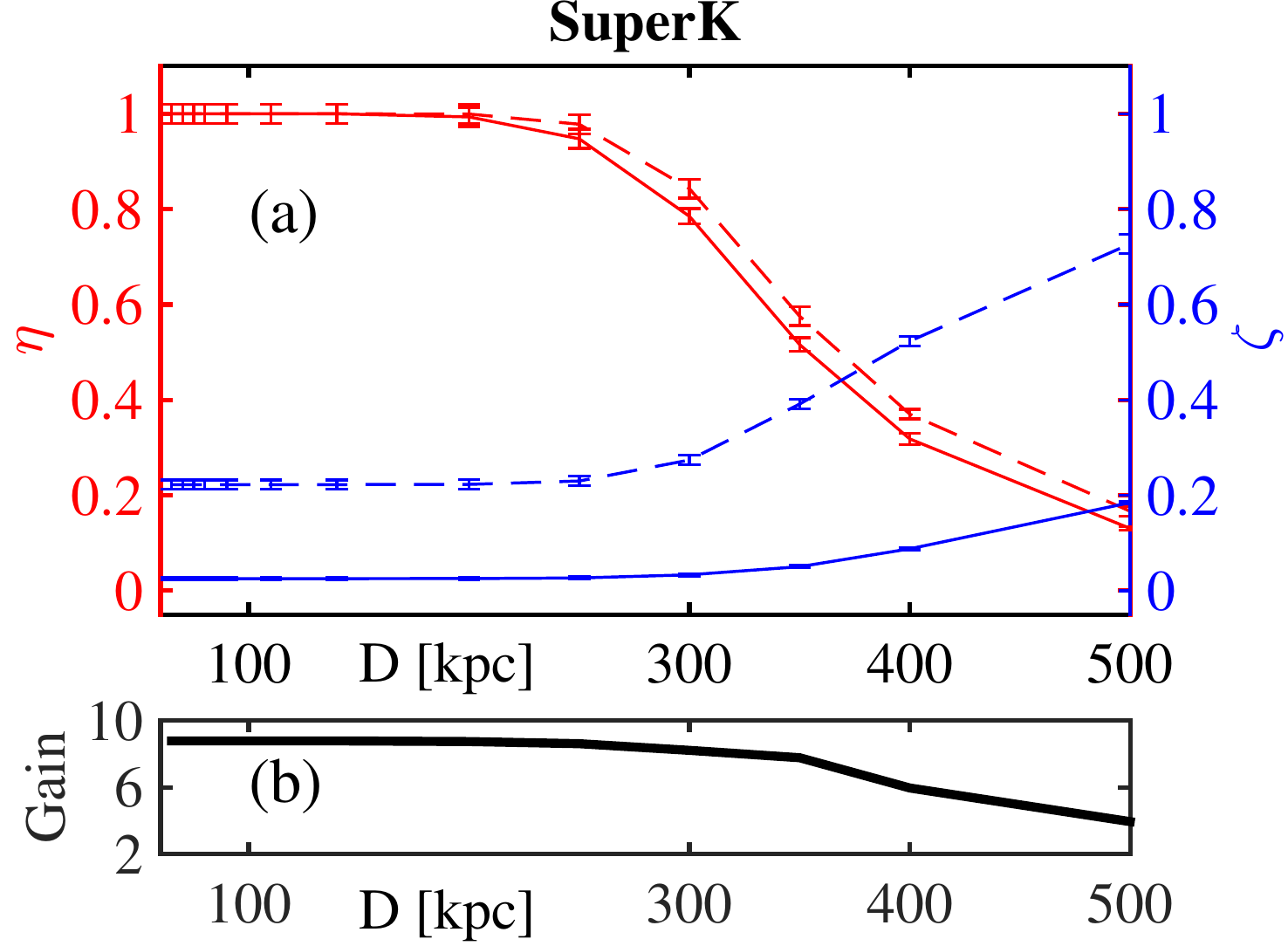}\\
$$
\caption{\em Panel (a): Red lines show the detection probability curves $\eta$ whereas blue 
lines show the misidentification probability curves $\zeta$ as a function of the source distance 
D and for SuperK detector working at $f^{im} \leq 1/\text{day}$. 
Solid (Dashed) lines are obtained by following the new proposed (standard) method for background reduction. 
Panel (b): The gain factor for SuperK, as defined in the text, versus the source distance D.     
\label{fig2}}
\end{figure}

As a leading example, we apply our search procedure to SuperK detector and 
we compare the detection probability obtained with the new procedure with  
the one achieved with the standard method. 
The selected statistical threshold $f^{im}\leq 1/\text{day}$ for SuperK 
is equivalent to a multiplicity cut of $m_i\geq 4$: this implies that an 
astrophysical burst, described as in our model, can be observed 
on average till a distance of $D_{sk}\simeq400$ kpc. This distance represents
the expected horizon for the detector operating at this statistical threshold. 

To show the improvement provided by our method we plot in 
Fig.\ref{fig2}(a) the detection efficiencies $\eta$ with red lines 
and the misidentification probabilities $\zeta$ with blue lines. 
In particular, dashed lines are obtained by using the standard procedure 
only based on statistical cut as described above, 
whereas the solid lines are obtained by applying our additional selection criterium. 
It is evident from the figure that the efficiency is unchanged 
till a distance of $\sim 200$ kpc, whereas the misidentification, $\simeq 23\%$ by 
using the standard procedure, drops to very small value  
$\simeq 3\%$ by applying the $\bar{\xi}$ cut.   

\begin{table*}[th]
\begin{center}
\begin{tabular}{ | l || c |c| c| c| c| c|}
		\hline
		\textbf{Detector} & M(kton) & $E_{thr}$(MeV) & $f_{bkg}$ (Hz) & $\bar{\xi} (\text{Hz})$ & $\bar{D}$(kpc) & G \\ \hline\hline
		Borexino & 0.3 & 1 &$0.048$ & 0.65 & 20 & 6.9 \\ \hline
		SuperK & 22.5 & 7 &$0.012$ & 0.72 & 200 & 8.9 \\ \hline
		KamLAND & 1 & 1 &$0.015$ & 0.77 & 50 & 13.4 \\ \hline
		LVD &  1 & 10  &$0.028$ & 0.72 & 40 & 14.0 \\ \hline
\end{tabular}
\caption {We show the considered detector features: the sensitive mass in kton (first column); the energy threshold used
for the analysis (second column); the average background frequency\cite{Abe:2016waf,Agafonova:2014leu,2009BorexinoColl,2003KamlandColl}(third column); the $\bar{\xi}$ value maximising the signal to noise ratio (fourth column); the largest distance $\bar{D}$ achievable without efficiency 
loss (fifth column); the gain factor obtained by using the new proposed method (last column).}
\label{tab:freq}
\end{center}	
\end{table*} 

A gain factor of the order of 10 implies that SuperK can work at a $f^{im}$ threshold  
10 times higher then the one based on the standard method
obtaining the same background reduction thanks to the new $\bar{\xi}$ cut. 
A detector working at an higher statistical threshold 
is sensitive to larger distances, so that our result can be also expressed in term of an increased horizon. 
Let us consider, for example, the threshold assumed by SuperK for normal warning 
in the online search for Supernova bursts\cite{Abe:2016waf}: $m_i \geq 25$, 
that corresponds to $f^{im}\leq (3.5\cdot10^{-10})/\text{year}$. This normal warning is sent to
SNEWS and it corresponds to an horizon of $147$ kpc. By considering the new additional cut, based
on the $\xi$ parameter, the same background reduction can be obtained by operating at a threshold
$f^{im}\leq (3.5\cdot10^{-9})/\text{year}$ corresponding to $m_i\geq 11$ and increasing the horizon up to $221$ kpc.

The improvement of SuperK is representative of the new search method, on the other hand, 
for completeness, we also investigate the others detectors. In particular, we show in the last two columns 
of Tab.\ref{tab:freq} the maximal distance $\bar{D}$(kpc) for which the additional $\bar{\xi}$ cut provides no efficiency loss  
and the corresponding gain factor, $\text{G}=\zeta/{\zeta}'$, calculated for such a distance as the ratio 
between the misidentification probability before and after the $\bar{\xi}$ cut. 
This gain factor as a function of the source distance is reported in Fig.\ref{fig2}(b) for SuperK.

\begin{figure}[h!]
$$
\includegraphics[width=0.8\textwidth]{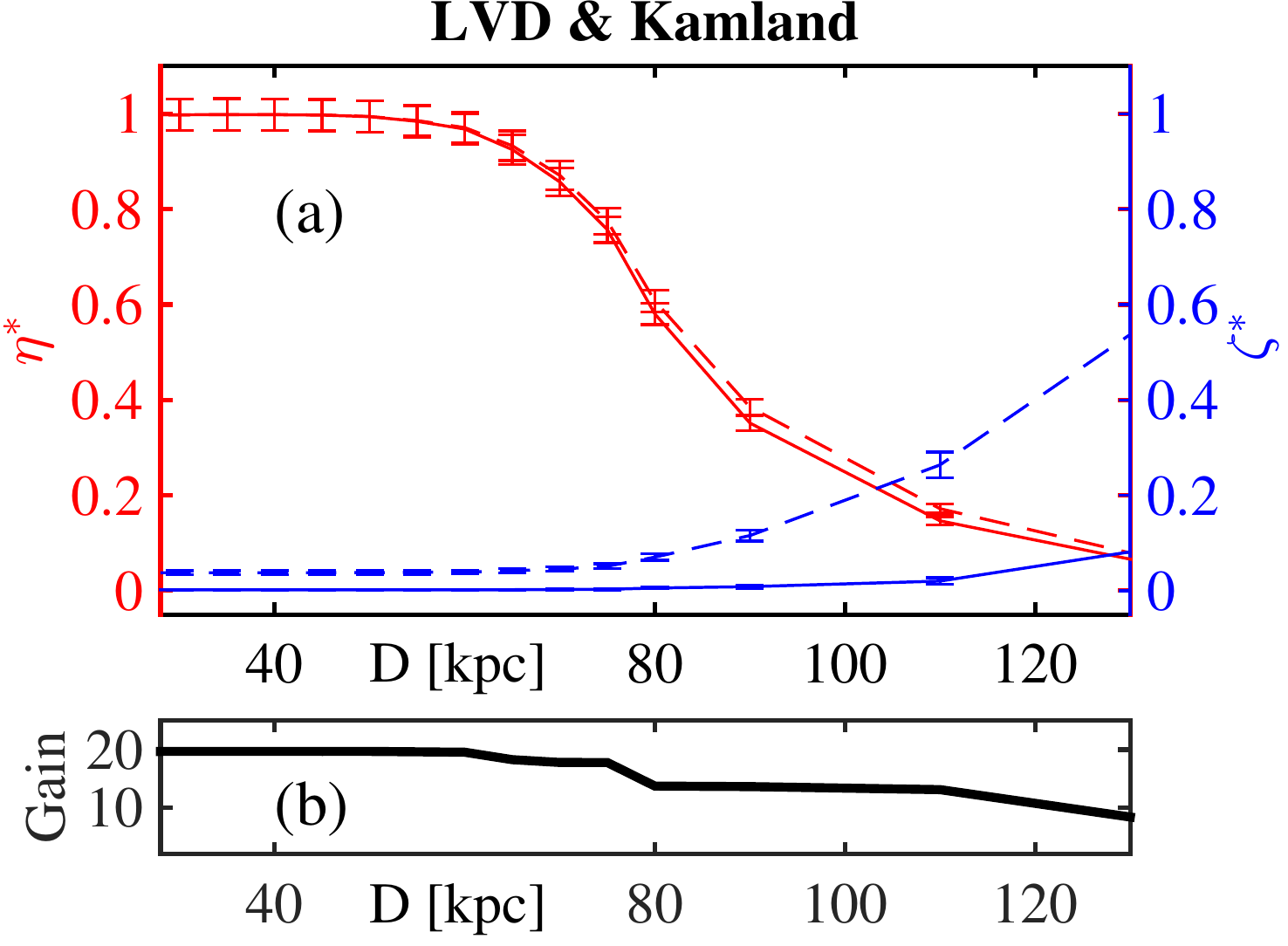}\\
$$
\caption{\em 
Panel (a): Red lines show the detection efficiencies curves 
$\eta^*$ whereas blue lines show the misidentification probability curves $\zeta^*$ for the network LVD $\&$ Kamland.
Solid (Dashed) lines are obtained by following the new (standard) method for background reduction. 
Panel (b): The gain factor for the network LVD $\&$ Kamland 
as a function of the source distance $D$. 
\label{fig3}}
\end{figure}

\section{Network extension.}  
In order to construct a list of candidate signal clusters, 
when two or more detectors operate together as a network, a further step is required, i.e. the temporal 
coincidence among clusters of different detectors within a time window 
that we assume to be $w_c=10$ seconds \cite{SNEWS}. 
The requirement to observe a coincident signal naturally decreases the background 
contamination, therefore increases the detection efficiency. 
In this case the statistical concept of false imitation frequency 
is substituted by the false alarm rate  
\begin{equation}
\text{FAR}=2w_c^{N_d-1}\prod_{X=1}^{N_d} {{f^{im}_{X}}},
\end{equation}
where $N_d$ is the number of detectors in the network and $f^{im}_{X}$ is the imitation frequency 
of each detector. 
Once that a threshold for the global FAR is defined, the corresponding threshold in $f^{im}_X$ 
for the different detectors depends on the network configuration.
Assuming for example as global FAR of the network $\le 1/\text{day}$, the required 
single detector threshold is $f^{im}_{X}\leq 66/\text{day}$ for a $2$-fold coincidence (i.e. two detectors network), 
whereas becomes $f^{im}_{X}\leq 265/\text{day}$ for the $3$-fold coincidence and so on.

Simulated astrophysical signals are injected inside the detectors background 
by taking into account the time of flight between the detectors.
The procedure for the clusters definition is the same as described in 
the previous section and, once that clusters are defined, only clusters coincident 
in the time window $w_c$ are selected.
The new additional cut based on the $\xi$ parameter is now applied to coincidences,  
we require that the product of the $\xi_X$ of coincident clusters is greater then the global cut value:
\begin{equation}
\overline{\xi}^*=\prod_{X=1}^{N_d}  \overline{\xi}_{X}.
\label{globalxi}
\end{equation}
The sensitivity of the neutrino network can be obtained by using 
an extended definition of the detection efficiency $\eta^*$, i.e. the number 
of astrophysical clusters surviving the statistical cut on $f_{im}^X$ that are found 
in temporal coincidence and are also characterised 
by a global $\overline{\xi}^*$ greater than the cut value defined in Eq.\ref{globalxi} 
over the total injected signals.
In a similar manner the network definition 
of the misidentification probability $\zeta^*$ becomes 
the ratio among background coincidences and the total number of found coincidences.

As a leading example, we show the case of LVD \& Kamland working together at a $FAR\le 1/\text{day}$. 
The detection efficiency and the misidentification probability of this network are showed 
in Fig.\ref{fig3}(a). As in the previous plot dashed lines represent 
the old method based on statistical cut plus temporal coincidence search, whereas 
solid lines show the same quantities obtained by adding the $\overline{\xi}^*$ cut as described above.
In particular, the misidentification probability is nearly constant until $75$ kpc around a value of $4\%$ with the standard procedure and decreases 
to a value around $0.2\%$ with the new cut. The gain factor obtained in this distance range is around $\sim 20$ as reported in Fig.\ref{fig3}(b).
This reduction of the misidentification can be also converted in term of a reduction of the FAR for the network. In other words, the network 
LVD \& Kamland operating at a FAR of $0.001/\text{day}  (0.365/\text{year})$ with the inclusion of our method, based on the $\xi^*$ cut, 
can reach the same background level of LVD \&Kamland working at the SNEWS threshold of $1/1000$years where only the statistical selection is applied. 

We apply this extended procedure to all the possible sub-networks of detectors 
including LVD, Borexino, Kamland and SK. In any case the improvement obtained 
is of the same order, indeed also for the cases of combined search between 
LVD \& Borexino or Kamland \& Borexino the gain factor obtained is of $\sim19$, 
however with a reduced distance $\bar{D}\sim 50$ kpc due to the lower 
sensitivity of Borexino.

\section{Discussion}
In conclusion, we propose a novel search method 
for astrophysical bursts of low-energy neutrinos. 
This method allows us a powerful discrimination between 
background and signal by exploiting their 
different cumulative rate. The achieved results show a 
decrease of the misidentification probability of a factor 10-20 
without loosing on detection efficiency.       

The proposed method can be applied both on single detector search and
in combined search among different detectors and can be easily implemented 
in the SNEWS online search for enhancing its detection potential and horizon.
Finally, we stress that the proposed method works 
for any low-energy neutrinos detector, water Cherenkov, liquid scintillator 
or argon based, that will be on data-taking in the next future, viz.
HyperK\cite{Abe:2016ero}, JUNO\cite{An:2015jdp} and DUNE\cite{Acciarri:2016ooe}.

Moreover combined search of core collapse supernovae 
with low-energy neutrinos and gravitational waves 
that are on going\cite{Leonor:2010yp,gwnu,Gromov:2017ncq}, can profit 
from this new procedure, being already based both on a  
complete data-sharing and on a combined search with different detectors.
This will also increase the detection probability 
for gravitational wave bursts expected from 
the here considered astrophysical sources.

\end{document}